\documentstyle[aps,preprint]{revtex}

\begin{document}
\draft
\preprint{{IMSc-98/18, IC/98/99.}}
\title{Emission of Fermions from BTZ Black Holes}
\author{ Arundhati Dasgupta\footnote{E-mail:dasgupta@imsc.ernet.in}}
\address{ The Institute of Mathematical Sciences,\\ 
            C. I. T. Campus, Chennai-600113, India,\\
          and\\
          Abdus Salam International Centre for Theoretical Physics,\\ 
           Strada Costiera, 11- 34014, Trieste, Italy.}
\maketitle
\begin{abstract}
The emission rate of fermions from 2+1 dimensional BTZ black
holes is shown to have a form which can be reproduced from
a conformal field theory at finite temperature. The rate obtained for
fermions is identical to the rate of
non-minimally coupled fermions emitted from  
a five dimensional black hole, whose near horizon
geometry is $BTZ \times M$, where M is a compact manifold.

\end{abstract}
\newpage
\section{Introduction} 
Understanding Hawking radiation from black holes using a unitary microscopic
theory has been a outstanding and difficult problem to solve. Nevertheless
the progress made in last few years suggests that resolution
of the problem may be near. In a series of papers, it has been shown that 
the Hawking
emission rates of a number of four and five dimensional black holes
is reproducible from a 1+1 conformal field theory
at finite temperature {\cite{haw,kleb,gub,das,hosh}}. In case of the five dimensional black
hole which is a solution of Type II B supergravity obtained by
wrapping  D5-branes and
D1-branes on $T^4 \times S^1$, it is known that the entire configuration can be
replaced by a long effective string along $S^1$ which 
is described by a
conformal field theory at finite temperature. However for
other black holes like the rotating ones, the origin
of the effective string is not well understood.
Recently, new light has been shed regarding the origin of
the underlying conformal field theory. It has been shown that
the above black holes can be mapped to the asymptotically anti-de
Sitter 2+1 dimensional BTZ black hole \cite{btz}
 times a compact manifold
\cite{sfe,malda}. In particular,
{\cite{malda,vij}} the near horizon
geometries of these black holes have the form BTZ $\times$ M, (or $AdS_3$ $\times
$ M with global identifications in case of rotating black holes \cite{cvet}) where
M is a compact manifold.  
The underlying microscopic theory of the BTZ black hole 
is known to be a 1+1 dimensional conformal
field theory \cite{carlip,hen,strom}.
Hence the conjecture that near horizon geometry
explains thermodynamics of black holes, and
the conformal field theory used to describe 
the BTZ $\times$ M geometry
gives the microscopic theories
of higher dimensional black holes.
The conjecture is supported by the evidences that the near horizon 
BTZ black hole obtained has the same entropy as the higher dimensional black hole,
and possesses the same decay rate for 
scalars as higher dimensional ones \cite{sen}. 

The 2+1 dimensional BTZ black hole is an interesting system
to study by itself. 
The black hole is asymptotically $AdS_3$, and has two horizons
with an ergosphere. For non-extreme BTZ black hole, there
is a non-zero Hawking temperature, and an observer outside
the outer horizon will detect
thermal radiation. However, since the metric is
asymptotically anti-de Sitter, the local temperature
measured by any time like observer decreases with distance
at spatial infinity. The situation is different from
flat space-time where temperature is constant at infinity.
Moreover in flat space-time, 
the asymptotic observer measures a decay rate
which is modified by the absorption coefficient
or the greybody factor of the black hole. The greybody
factor is defined as the ratio of the total number of particles 
entering the 
 horizon and the incoming flux at spatial infinity.
For asymptotically anti-de Sitter black holes, this is a little difficult 
to study, as 
spatial infinity for these black holes 
constitute a time like
surface through which information can enter and leave. The usual way
to deal with this is to impose boundary conditions on fields such that the surface acts like a reflecting
wall.
In that case, the black hole is in equilibrium with thermal radiation and there
is no net flow of flux 
across any time like surfaces and the concept
of greybody factor remains obscure.
Nevertheless, 
an attempt can be made to define the Hawking rate
in the same way as asymptotically flat space-times.
Using this in {\cite{sen,lee}}, it was shown that the decay rate
for scalars has the same form as the higher dimensional
black holes, and can be reproduced from a 
conformal field theory at finite temperature.

In this paper we address the issue of fermion emission from BTZ
black holes. This is interesting to investigate by itself, as
it is yet to be checked that the above stated result for
scalars applies universally to all particle emissions from
the BTZ black hole. 
Not only that, the result constitutes important evidence for the conjecture that
near horizon geometry of higher dimensional black holes
encode information about it's thermodynamics. To define the Hawking emission 
rate for the BTZ black hole, we do not take an asymptotic
observer, but an observer stationed at a radial distance
$\rho\sim l\gg \rho_+$, where $l$ is related to the cosmological constant.
A motivation for this is that at this position, the observer in
the BTZ geometry measures a local temperature  
 equal to the Hawking temperature of the black hole.
This issue is discussed in section III. Also, for purposes of the
greybody calculation, we assume that there is a flow
of flux into the black hole. 
This is
perhaps a relevant physical situation to consider if we want
to answer questions about higher dimensional black holes as
flux can flow into ``near horizon geometry'' (BTZ). 
Using the above assumptions, we show that
the fermions in the BTZ geometry have exactly
the same form of the Hawking emission rate as the non-minimally
coupled fermions in five-dimensional black hole \cite{hosh}. Indeed this
strengthens the conjecture. The form obtained can be reproduced
from a conformal field theory at finite temperature. 
Since a
satisfactory derivation of higher dimensional fermion
decay rates does not exist from the effective string picture \cite{das,hosh},
this result is very useful.
A microscopic derivation of the rate found here by using the conformal field theory
describing the BTZ black hole will finally clear the
issue.
Moreover, three dimensional fermions are
much easier to handle than the higher dimensional ones,
and this calculation can be used to predict decay rates for higher
dimensional fermions. In particular, 
the 4 dimensional black hole of M-theory also has BTZ as it's near horizon
geometry and,
it is interesting to predict the nature of the fermions of 11-D
supergravity compactified to 4-dimensions,
 which have the same decay rate as the 
fermions considered in three dimensions. 

In the next section, we solve the equation of motion of the fermion
in the BTZ background. We find that the equation of motion is
exactly solvable, and yields a hypergeometric equation. We can
choose the ingoing solution at the horizon, and determine the
flux which flows down the hole. It should be mentioned that taking
minimally coupled fermions instead of those considered here,
does not give a meaningful result. In the following section,
we derive the greybody factor. We determine the solutions
in the region $\rho\sim l$ and determine the flux which
flows towards the black hole horizon. 
Using this, we can calculate the greybody factor for the black hole,
and hence the decay rate.
In the third section, we give a comparison of the rate obtained
here, and that obtained for 
higher dimensional black holes. In the last section,
we include a brief discussion.
\section{Equation of motion of the fermion in BTZ background}
Einstein's gravity in 2+1 dimensions is essentially
topological, and does not admit black hole solutions. However,
gravity with a negative cosmological constant, has non-trivial
solutions. The action for this is:
\begin{equation}
S= \frac{1}{2\pi}\int d^3 x \left( R + 2l^{-2}\right)
\end{equation}
Where $R$ is the scalar curvature and $\Lambda = -(1/l^2)$ is the cosmological constant;
$G=1/8$ according to the conventions of \cite{btz}.
The space with constant negative curvature, is called
anti-de Sitter space. This space is invariant under SO(2,2)
group, which is larger than the usual Poincare group of flat space time.
The appropriate 
covariant derivative for spin- half fields is:
\begin{equation}
{D}=
\gamma^{\nu}\left(\partial_{\nu} + \omega_{\nu} + g~e_{\nu a}\gamma^{a}\right)  
\label{cov}
\end{equation}
Where, $\omega_{\nu}$ is the spin connection, $e_{\nu a}$ is the triad 
and $g=1/2l$ is related to the cosmological constant. 
The BTZ metric is derived 
by appropriate identifications of anti-de Sitter space time, and it's asymptotic properties
are the same as anti-de Sitter space  
\cite{btz}. Hence the above covariant derivative is relevant for our
purposes. To write down the fermion equation in the BTZ background,
we study the metric first. The BTZ metric in coordinates $\rho, t$ and $\phi$ is
($0<\rho<\infty, 0<\phi<2\pi$):
\begin{equation}
ds^2= - \frac{\Delta^2}{l^2 \rho^2} dt^2 + \frac{l^2
\rho^2}{\Delta^2} dr^2 + \rho^2\left(d\phi -
\frac{\rho_+\rho_-}{l \rho^2} dt\right)^2,
\end{equation}
\begin{equation}
\Delta^2 = {\left( \rho^2 -\rho_+^2\right)\left(\rho^2 -
\rho_-^2\right)}. 
\end{equation}
Clearly, the metric represents a rotating black hole, with two horizons at $\rho_+$ and
$\rho_-$. The angular momentum of the black hole is $J=2\rho_+\rho_-/l$, and it's
mass is $M=(\rho_+^2 + \rho_-^2)/l^2$. The metric can be written in more convenient coordinates,
with the radial coordinate $\rho$ defined in terms of hyperbolic coordinate $\mu$.
The redefinition is:
\begin{equation}
\rho^2=\rho_+^2 \cosh^2{\mu} - \rho_-^2\sinh^2\mu.
\end{equation}
The spatial infinity corresponds to $\tanh\mu\rightarrow 1$. In this coordinate, the
metric takes a form:
\begin{equation}
d s^2= -\sinh^2{\mu}\left( \rho_+ \frac{dt}{l} - \rho_ -d\phi \right)^2 + l^2 d\mu^2 + \cosh^2\mu
\left(-\rho_-\frac{dt}{l} + \rho_+d\phi\right)^2
\end{equation}
A convenient set of linear combinations of $t$ and $\phi$  gives us
$x^+=\rho_+ t/l - \rho_-\phi$ and $x^-=-\rho_-t/l + \rho_+\phi$, and the killing directions
of the metric are $\partial_{x^+}$ and $\partial_{x^-}$. The triads are chosen 
in an appropriate local Lorentz frame
in the tangent plane.

\begin{eqnarray}
e_{x^+}^{0}&= & \sinh\mu\;\;\; e_{x^-}^2= \cosh\mu \nonumber \\
e_{\mu}^{1}&=& l.
\end{eqnarray}
The non zero spin connections for this are:
\begin{eqnarray}
\omega_{x^+}& =&-\frac1{2l}\cosh\mu\;\sigma^{0 1} \nonumber \\ 
\omega_{x^-}& =&\frac1{2l}\sinh{\mu}\;\sigma^{2 1},
\end{eqnarray}
here $\sigma^{a b}=1/2[\gamma^a,\gamma^b]$.
Using the above, we substitute them in (\ref{cov}) and obtain the fermion equation on the BTZ space-time as
\begin{equation}
\gamma^1\frac1l\left(\partial_\mu + \frac{\sinh\mu}{2\cosh\mu} + \frac{\cosh\mu}{2\sinh\mu}\right)\psi
+ \gamma^0\frac{\partial_{x^+}\psi}{\sinh\mu} + \gamma^2\frac{\partial_{x^-}\psi}{\cosh\mu} +\frac{1}{2l}\psi =0
\end{equation}
We take the representation of gamma matrices to be:
$\gamma^0=\imath\sigma^2, \gamma^1=\sigma^1, \gamma^3=\sigma^3$. 
The killing isometry requires that $\partial_{x^{\pm}}\psi= -\imath k^{\pm}\psi$, where $k^{\pm}$ are constants
depending on  the energy and azimuthal eigenvalues $\omega$ and
$m$ respectively. In fact they can be determined, and
$k^+=\left( \omega - m\Omega\right)/(2\pi l~
T_H)$ and $k^-=\left(\rho_-\omega -\rho_+m/l\right)/(2\pi l\rho_+  T_H)$. The Hawking temperature is $T_H= (\rho_+^2 - \rho_-^2)/2\pi l^2 \rho_+$ and $\Omega=
J/2\rho_+^2$. We take the following form for the wavefunction:
$$\psi= \frac{e^{-\imath(k^+x^+ + k^- x^-)}}{\sqrt{\sinh\mu\cosh\mu}}\pmatrix{\psi_1\cr
\psi_2\cr}$$
The radial equations for the two components are determined as

\begin{eqnarray}
\left(d_{\mu} - \frac{\imath l k^+}{\sinh\mu}\right)\psi_2&=&-\left(\frac12 - \frac{\imath l k^-}{\cosh\mu}\right)\psi_1\\
\left(d_{\mu} + \frac{\imath l k^+}{\sinh\mu}\right)\psi_1&=&-\left(\frac12 + \frac{\imath l k^-}{\cosh\mu}\right)\psi_2
\end{eqnarray}
Interestingly, to separate the wave functions, we have to go to a different basis
of wave functions. Let us call them $\psi'_1$ and $\psi'_2$, defined as,
\begin{eqnarray}
\psi_1 + \psi_2& =&\left(1- \tanh^2\mu\right)^{-1/4} \sqrt{ 1 + \tanh\mu}\left(\psi'_1 + \psi'_2\right)\\
\psi_1 - \psi_2& = &\left(1- \tanh^2\mu\right)^{-1/4}\sqrt{ 1 - \tanh\mu}\left(\psi'_1 - \psi'_2\right)
\end{eqnarray}
We obtain the 
equations in coordinates $y=\tanh\mu$ for the $\psi'$ 
below as:
\begin{eqnarray}
(1-y^2)d_y\psi'_2 -\imath l \left( \frac{k^+}{y} +k^- y\right) \psi'_2 &= -&\left\{ 1 -\imath l (k^+ +k^-)\right\}\psi'_1\\
(1-y^2)d_y\psi'_1 +\imath l \left( \frac{k^+}{y} +k^- y\right) \psi'_1 &= -&\left\{ 1 +\imath l (k^+ +k^-)\right\}\psi'_2
\label{fine}
\end{eqnarray}
The equations are now very easily separable. The second order differential
equation obtained from the above two equations can be cast in a simple form in the variable $y^2$ which we denote as $z$ for convenience.
\begin{equation}
z(1-z)d^2\psi'_1 + \frac12\left( 1 -3z\right)d\psi'_1 + \frac14\left(\frac{-\imath lk^+ + l^2 k^{+2}}{z} + \imath l k^- - l^2 k^{-2} -\frac{1}{1-z}\right)\psi'_1=0
\label{hyp}
\end{equation}
The solution to this equation is determined to be $\psi'_1= z^m(1-z)^n F(\alpha,\beta;\gamma; z)$, where $F$ is a hypergeometric function. For the ingoing function, the constants
are as follows: $m = 1/2 + \imath l k^+/2, n= -1/2$, and the hypergeometric
parameters are: $\alpha= \imath l (k^+ + 
 k^-)/2 + 1/2, \beta = \imath l (k^+ - k^-)/2  $
and $\gamma = \imath l k^+ + 3/2$.  The ingoing function is so chosen that
at the horizon, the wave function has the dependence $\psi\sim exp(\imath
(\omega/4\pi T_H )\log z)$. 
The solution for the other component of the wave function can be determined
easily now. It is: $\psi_2= z^{ilk^+/2}(1-z)^{-1/2} \left(-(\gamma -1)/\alpha\right) F(\alpha -1, \beta; \gamma -1;z)$, where $\alpha, \beta, \gamma$ are constants as defined above. The flux for this function as shown in the next section is negative,
indicating a flow into the black hole.

Thus we see that the fermion equation of motion in the BTZ background
is exactly solvable. It is interesting to note that $n=0$ corresponds
to a minimally coupled fermion, and in that case $\gamma=\alpha +\beta$. 
The hypergeometric solution is not well behaved, and does not
converge as $z\rightarrow 1$. There can be other kinds of couplings
to the BTZ metric, and they will be interesting to investigate. 
The BTZ black hole is locally anti- de Sitter
space, with global identifications. 
It will be interesting to see
whether the solutions obtained here can be related to those
obtained for $AdS_3$ in {\cite {hyun1}}, modulo the global identifications.
\section{Grey Body Factor}
The black hole grey body factor is also the absorption coefficient of the black hole. The geometry of the black hole provides a kind of potential barrier for the 
fields propagating on it. Only a fraction of the incoming flux at infinity is
absorbed by the body, and rest is reflected back. In order to determine the total 
Hawking radiation rate of the observer, sitting far away from the black hole,
we need to calculate this absorption rate. Indeed, as in ordinary
quantum mechanics, the black hole absorption rate, which we denote
by $\sigma_{\mbox{abs}}$ is related to the ratio of the ingoing flux at horizon
and incoming flux at infinity. The fermion flux into the horizon
will be determined by the current which enters the horizon. Usually,
to probe the black hole geometry, an incoming plane wave is
taken at infinity.
However, here, for our purposes, we take the incoming flux in the
region $\rho\sim l\gg\rho_+$ as the incident flux on the black hole. This
would correspond to an BTZ observer, sitting at finite $\rho$,
detecting radiation. Though, the physics of this picture is not very clear
, there are a number of reasons for choosing this. 
In curved space-time, an observer
measures a thermal spectrum depending upon his local temperature, which is
$T_H/ \sqrt{g_{00}}$. In asymptotically flat space time, $ \sqrt{g_{00}}
\rightarrow 1$ as $\rho\rightarrow \infty$. However, this is not the
case in asymptotically anti-de Sitter space-time where $\sqrt{g_{00}}\sim \rho$
at spatial infinity. For small mass
BTZ black holes, i.e. $\rho_+\ll l$, it is easily seen that,
$\sqrt{g_{00}}\rightarrow 1$ when $\rho\sim l$. This motivates the choice
of the observer. Moreover, to compare
our final answer with higher dimensional black hole rates, going infinitely
away from the horizon would imply a modification of the near horizon BTZ geometry,
and we are not interested in probing that region. 
As $\rho\sim l$, the black hole metric is same as asymptotically anti-de Sitter space.  
Solutions determined in this metric is also the same as that obtained
in the vacuum solution of the black hole \cite{btz2}. The metric is
($\rho\gg\rho_+$):
\begin{equation}
ds^2= -\frac{\rho^2}{l^2} dt^2 + \frac{l^2}{\rho^2} d\rho^2 + \rho^2 d\phi^2
\end{equation}
To, determine the wave functions we then solve the radial equations:
\begin{equation}
\left(\rho~\partial_{\rho} \pm \frac{\imath \omega l^2}{\rho}\right)\psi^f_{\;\; 1 (2)} = -\left(\frac12 \pm \frac{\imath m}{\rho}\right)\psi^f_{\;\;2(1)}
\end{equation}
To separate this set, we go to a frame in which, $\psi'^f_{\;\;1}= \psi^f_{\;\;1} + \psi^f_{\;\;2}$
and $\psi'^f_{\;\;2}=\psi^f_{\;\;1} -\psi^f_{\;\;2}$. The equation can be exactly solved in this
frame. The solutions are determined, in terms of Bessel functions,
\begin{eqnarray}
\psi'^f_{\;\;1} &= &{\sqrt x}\left( A _{1}J_0(\Lambda x) + \imath A_2 N_0(\Lambda x)\right)\\
\psi'^f_{\;\;2}&=& \frac{\imath \sqrt x}{E}\left( A_1J_1(\Lambda x) + \imath A_2 N_1(\Lambda x)\right)
\label{was}
\end{eqnarray} 
Where $J_{n}$ and $N_{n}$ are bessel functions of the first
and second kind. $A_1$ and $A_2$ are arbitrary constants
of integration. Also $x= 1/\rho, \Lambda=
l\sqrt{\omega^2l^2 - m^2}, E= l(\omega l +m)/\Lambda$. Note that the
same solutions will survive when $\rho\rightarrow \infty$ in anti-de Sitter
space.
The interesting aspect about anti-de Sitter space is that $\rho\rightarrow\infty$
is a time like surface. Hence, it is necessary to specify boundary conditions,
which are either Dirichlet or Nueman on the surface. These boundary
conditions, also called reflective boundary conditions {\cite{avis}} can
be realised in the set of functions defined in ({\ref{was}}). On choosing 
$A_2 =0$, this condition can be ensured for the above wavefunctions.
It is easy to check that in that case $\sqrt{\rho}\psi=0$ for $\rho\rightarrow \infty$. However, here we are
not interested in making the wall totally impervious. Instead, we 
are forced to take $A_2$ have non-zero values if we want the wavefunction
to match the wavefunction determined in
({\ref{hyp}}) continued to $z\rightarrow 1$. This shows that, our choice of a net inflow
of flux into the black hole ensures that we donot have reflecting
boundary conditions at infinity. For asymptotic anti- de Sitter space, this might be related
to the transparent boundary conditions defined in {\cite{avis}}.
Before we can determine the greybody factor using this far solution,
we need to match this with the ingoing wavefunction at the horizon
since we want the wavefunction to be continuous in space-time. 
To do that, we continue the solution of ({\ref{hyp}})
to $z\rightarrow 1$ \cite{bateman}. Now, with the scalings and redefinitions, the
radial wavefunction is:
\begin{eqnarray}
\psi_{1}&\rightarrow &\frac{E_1\sqrt{N (\rho_+ + \rho_-) }}{ (2 \rho)^{3/2}}\left\{ -\left(1 - 2\Psi(\alpha)\beta - 2\Psi(\beta +1 )\beta\right) + 4\beta\left( \log \left(\frac{\sqrt N}{\rho}\right) + C\right)\right\}\\
\psi_{2}&\rightarrow& \frac{\sqrt{\rho_+ - \rho_-}}{\sqrt{2 \rho}} E_1
\end{eqnarray}
Where, 
\begin{equation}
E_1= \left(\frac{\Gamma(\gamma)\Gamma(\gamma-\alpha-\beta)}{\Gamma(\gamma -\alpha)\Gamma(\gamma - \beta)}\right),
\end{equation}
 $\Psi$ are the digamma function (subsequently, we write $ 2\beta(\Psi(\alpha) + \Psi(\beta + 1))\equiv \Psi(\alpha,\beta)$). Also $ N=\rho_+^2 - \rho_-^2$, and $\Psi(1)=-C$ (euler's constant).
The factors $\sqrt{(\rho_+ \pm \rho_-)/N}$ enter as this wavefunction given
in terms of $t,\rho,\phi$ coordinates is lorentz rotated from the
wavefunction obtained in $x^+,z,x^-$ coordinates. 
Note in the above, we have taken, $\sqrt{N}/\rho \ll 1$ or in other words, $\sqrt N \ll
l$.
The wavefunctions obtained in (\ref{was}) , can be cast in a similar form when $\Lambda/\rho
\ll 1$. (For $m=0$, this indicates that $\omega l\ll 1$)
\begin{eqnarray}
\psi^f_{\;\;1}&\approx& {\frac{1}{\rho^{3/2}}}\left( A_1 + \frac{2 \imath A_2}{\pi}\left( \log(\frac{\Lambda}{\rho}) + C \right) \right) \\
\psi^f_{\;\;2}&\approx& \frac{2 A_2 }{\pi E \Lambda \rho^{1/2}}.
\end{eqnarray}
The asymptotic constants are determined from the above equations, 
\begin{equation}
A_1= -\frac{\sqrt{N (\rho_+ + \rho_-)}}{2 \sqrt 2}E_1 \left(1 - \Psi(\alpha,\beta)\right) \;\; A_2 = \frac{\pi E \Lambda}{2\sqrt2}\sqrt {\rho_+ -\rho_-} E_1
\end{equation}
The fermionic flux is given by:
\begin{eqnarray}
{\cal F}= \sqrt {-g} J^{\rho}& = &\rho\; \bar{\psi}e^{\rho}_1\gamma^1 \psi\\
\end{eqnarray}
The incoming flux at $\rho=l$ is determined as (The flux obtained from the
above constants is multiplied by $l^2/N^2$ for normalisations) 
\begin{equation}
{\cal F}^f= -\frac{l}{ 8 N}|E_1|^2\left( 2 - 2 Re\Psi(\alpha,\beta)\right)
\end{equation}
The absorption coefficient is defined as the ratio of total
number of particles entering the horizon with the incoming 
flux at infinity \cite{unruh}. The total number of particles entering
the horizon is:
\begin{equation}
P= -\int \sqrt{-g} J^{\rho}\rho_+ d\phi
\end{equation}
This is equal to: $A_H l/4N |(\gamma -1)/\alpha|^2$, where $A_H$ is the area of the
horizon. The absorption coefficient
is then determined as (for $m=0) $:  
\begin{equation}
\sigma_{\mbox{abs}}= \frac{A_{H}}{1 - Re\Psi(\alpha,\beta)} \left(\left|\frac{\gamma-1}{\alpha}\right|^2\right)\left|\frac{\Gamma(\gamma)\Gamma(\gamma-\alpha-\beta)}{\Gamma(\gamma -\alpha)\Gamma(\gamma - \beta)}\right|^{-2}
\end{equation}
Here, Re$\Psi(\alpha,\beta)= -\left(\omega/4 T_L\right)\tanh\left(\omega/4T_R\right)
- O(\beta^2)$ for m=0. 
Using the expressions for the hypergeometric parameters as given in the
earlier section, the greybody factor can be written in a interesting
form: 
\begin{equation}
\sigma_{\mbox{abs}}= \frac{\omega A_{H}} {4 T_L( 1 + \omega/4 T_L \tanh(\omega/4 \pi T_R))}\frac{\exp(\omega/T_H) +1}{(\exp(\omega/2T_L)-1)(\exp(\omega/2T_R) + 1)}
\end{equation}
Where, the quantities $T_L$ and $T_R$ are defined by:
\begin{equation}
\frac1{T_L} = \frac1{T_H}\left( 1- \frac{\rho_-}{\rho_+}\right) \;\;\;\; \frac1{T_R} = \frac1{T_H}\left( 1 +\frac{\rho_-}{\rho_+}\right)
\label{abs}
\end{equation}
The extremal limit, defined 
by taking $\rho_+\rightarrow \rho_-$, corresponds to $T_L \gg T_R$.
Clearly, taking the above limit in equation (\ref{abs}), the absorption
coefficient reduces to $A_H/2$.
The Hawking radiation rate will be now a product of thermal distributions, instead of being a single fermionic distribution. 
Infact, it is:
\begin{equation}
\Gamma_H= \frac{\omega A_H }{4 T_L}\frac{ d^2k}{(exp(\omega/2T_L)-1)(exp(\omega/2T_R) + 1)}\end{equation}
This is precisely the form expected for emission rates from an underlying conformal theory
at finite temperature \cite{gub}. The fermion in the bulk couples to
operators of the 1+1 dimensional conformal field theory. The system is at finite temperature $T_H$, which can be
split into left and right temperatures such that $1/T_L + 1/T_R=2/T_H$. The decay
rate at finite temperature due to the coulping stated above is calculated to have the form of a product
of left and right distributions. 
The fermions are associated with rightmoving temperature, indicating
that fermions considered here couple to chiral conformal
operator. Using the results 
of {\cite{gub}} it can be predicted that the conformal
fermion couples to operator in the conformal theory of the form
$O^+ O^-$, where $O^+$ is rightmoving, and has conformal weight $1/2$
and $O^-$ is leftmoving with weight $1$. It will be interesting to
determine the nature of the coupling as that fixes the coefficients exactly.

\section{Comparison with Higher dimensional black holes}
One of the reasons behind the renewed interest in three dimensional
black holes, is the fact that near horizon geometries
of certain higher dimensional stringy black holes are BTZ times a 
compact manifold. Here we briefly review this mapping \cite{vij,mals}
and discuss the implications.
The solution due to RR charged one branes and five branes
wrapped on $T^4\times S^1$, and Kaluza Klein momenta along $S^1$
in 10 dimensions, has a near horizon geometry $BTZ \times T^4 \times S^3$.
The radius of the $S^3$ direction is $l=r_1 r_5$, where $r_1$ and $r_5$
are related to the one brane and five brane charges respectively.
The time, transverse radial distance $\rho$ and $S^1$ direction ($\phi$) constitute the BTZ black hole coordinates. 
The ordinary kaluza-klein reduction of the 10-D solution on $T^4\times S^1$ yields a 5-D
black hole, which preserves $N=8$ supergravity. The entropy of the 5-D black hole is equal to the entropy of the near horizon BTZ black hole
and scalar decay rate equals the decay rate for scalar emission from BTZ black holes.
Here we make a comparison for
fermion decay rates.
In \cite{hosh}, it has been shown that the SUGRA
fermions of $N=8$ supergravity, have a Hawking decay rate for the
five dimensional black hole as 
\begin{equation}
\Gamma^5_H= A^5_H \frac{\omega}{4 T_L}\frac{d^4k}{(exp(\omega/2T_L) -1)(exp(\omega/2T_R) +1)}
\label{high}
\end{equation}
Clearly, our decay rate is identical to this decay rate. The temperatures
of the left and right distributions are exactly the same as given in (\ref{abs})
and $A^5_H$ is the area of the horizon of the five dimensional black hole.
A interesting point to note is that the rates can be matched upto exact coefficients
if we choose to factor out the phase space factors of $S^3$ and $\phi= x_5/l$ ($x_5$ is the $S^1$ direction, with radius $R$) from the
decay rates, as $A_H^5/A_H^3=\pi l^3/R $. However, an observer in five dimensional
space detecting particles at infinity sees all the three
dimensions of $S^3$ as uncompactified. So, it is not clear what the above result implies.
However, it can be said that our result confirms the observation about scalar decay rates.
The range of frequency for 
both the calculations, $\omega r_1\ll 1$, is also same. 

The exact matching observed above provides a basis to predict rates for
non-minimally coupled fermions which propagate on the background
of the four dimensional N=4 SUGRA black hole obtained by compactifying M-theory (11-D
supergravity)
on $T^6\times S^1$. To identify
the required fermion one requires to take the equation of motion in the 11 D
Supergravity solution, and take the near horizon geometry limit
as described above. All fermions which will couple in the same
way as in equation (\ref{cov}) in 
the BTZ part, can then be predicted to have the rate obtained in this paper.
The metric in 11-Dimension is due to 3 M 5-branes
wrapped on $T^4 \times S^1$, i.e. directions $x_4..x_{11}$ and a boost in the $x_{11}$, ($S^1$) direction.
The near horizon 
limit results in the metric splitting up into a BTZ$\times S^2\times T^
6$, where $S^2$ is the two sphere of the noncompact t,r,$\theta, \phi$
dimensions of the four dimensional black hole. The radius of the two
sphere is $R=l/2= (r_1 r_2 r_3)^{1/3}$, where $r_i$ are related to the charges
of the black hole.
As in the five dimensional 
case here $\phi = x_{11}/R_{11}, \rho^2= 2 R_{11}^2(r_0 + r_0\sinh^2\sigma')$
($R_{11}$ is the radius of $x_{11}$), and time form the BTZ coordinates. To find the relevant fermions which
will have the rate as found in this paper, we start from the
11-D gravitino $\psi_M$. Clearly gravitino with vector polarisation along
$x_{11}$ or the other $x_1, x_2, x_3$ directions will not satisfy our requirements. We
take a representative $\psi_5$, as in the near horizon limit, all the
torus directions are similar, apart from constant scalings. In this 
limit, since $g_{ii}=$const, $i=4..9$, we can split the 11-D
equation of motion as:
\newcommand{\D}{D\!\!\!\!/}
\begin{equation}
\left({\D_3} + \frac2l{\D_{\Omega }}\right)\psi=0
\end{equation}
Where $\D_3$ and $\D_{\Omega}$ are the dirac operators in the BTZ and
the two sphere metrics respectively. To get simultanouos eigenstates
of both the operators, we multiply by the two dimensional chirality
matrix $\Gamma_2=\imath \Gamma_a\Gamma_b$. Thus for $\Gamma_2{\D_{\Omega }}\psi=\lambda\psi$, the equation of the fermion in the near horizon limit is:
\begin{equation}
\left({\D_{3 \mu}\;'} + \frac{2\lambda}{l}\right)\psi=0
\end{equation}
For $\lambda=1/4$, we have the required fermion ($\D\;'\equiv \Gamma_2\D$). It is not very difficult
to solve the eigenvalue equation stated above.
The arguement given here is hueristic, and we have not been careful
about the supersymmetry preserved by the background metric. It is to be
checked whether the fermion taken above falls in the $N=4$ multiplet,
as the four dimensional black hole preserves $N=4$ super symmetry.
However, it is an interesting calculation, and is under further investigation
at present.

\section{Discussions}
In this paper we have calculated emission rate of fermions from
BTZ geometry, using techniques of asymptotically flat space-time
calculations, like the greybody factor. However, since the
physical situation we are interested in is when BTZ occurs
as the near horizon geometry of higher dimensional black hole,
this is justified. We show that indeed the BTZ calculation reproduces the
rate of the non-minimally coupled fermions
in the background of a five dimensional black hole
whose near horizon geometry is $BTZ \times S^3$. The fact that the rate observed by a 
BTZ observer at $\rho\sim l$ looks identical
to that of an asymptotic observer in a five dimensional
black hole is interesting. The physical implications of this
are still not clear, but the answer might lie in the
location of the degrees of freedom of the underlying conformal field theory.
There are several ways to approach the problem. 
It is known that
2+1 gravity can be cast in the form of Chern Simon theory, which induces
a conformal field theory on the boundary. However, on inclusion of matter
fields the theory is no longer topological, and the same conclusions
cannot be drawn about the entropy. Hence, it is not clear how to
study Hawking emission in the above frame work. Recently, matter fields have been treated as a classical perturbation in the
Chern Simons action, and the decay rate obtained for scalars{\cite{sachs}}. The agreement
with the black hole decay rate is remarkable, and calls for further investigation. Apart
from this, the BTZ black hole is asymptotically anti de Sitter, and has
a conformal field theory living on it's boundary \cite{hen,strom}.
With the AdS/CFT correspondence, it is known now, that string theory on orbifolds of
$AdS_3$ times a compact manifold $M$ is dual to a super conformal field theory 
whose target space is symmetric product of $M$ \cite{mals}. In this matter fields are automatically included, and it
will be interesting to study the decay rates, using this approach. 
\noindent
\begin{center}
{\bf ACKNOWLEDGEMENTS}
\end{center}
We are grateful to P. Majumdar, D. Page, A. Sen, S. Das, P. Durganandini and T. Sarkar
for useful discussions. We thank Theory Group CERN for hospitality
where a part of the work was done.


\begin{references}
\bibitem{haw}C. Callan and J. M. Maldacena, Nucl. Phys. {\bf B472} (1996)591;
A. Dhar, G. Mandal and S. R. Wadia Phys. Lett. {\bf B388} (1996) 51;
S. R. Das and S. D. Mathur, Nucl. Phys. {\bf B478} (1996) 561;
J. Maldacena and A. Strominger, Phys. Rev. {\bf D55} (1997) 861;
Phys. Rev. {\bf D56} (1997) 4975; M. Cvetic and F. Larsen Nucl. Phys. {\bf B
506}, 107.
\bibitem{kleb} S. S. Gubser and I. R. Klebanov, Phys. Rev. Lett 77 (1996)
4491.
\bibitem{gub} S. S. Gubser, Phys. Rev. {\bf D56} (1997) 4984.
\bibitem{das}S. Das, A. Dasgupta, P. Majumdar and T. Sarkar, hep-th/9707124.
\bibitem{hosh} K. Hosomichi, Nucl. Phys. {\bf B524} (1998) 312.
\bibitem{btz}M. Banados, C. Teitelboim, J. Zanelli, Phys. Rev. Lett. 69(1992) 1849.
\bibitem{sfe} S. Hyun, hep-th/9704005, K. Sfetsos and K. Skenderis, Nucl. Phys. {\bf B517}(1998) 179.
\bibitem{malda}J. Maldacena, hepth/9711200.
\bibitem{vij} V. Balasubramanian and F. Larsen, hep-th/9802198.
\bibitem{cvet} M. Cvetic and F. Larsen, hep-th/9805097.
\bibitem{carlip}S. Carlip hepth/9806026 and references therein.
\bibitem{hen} J. D. Brown and M. Henneaux, Comm. Math. Phys. {\bf 104} (1986),207.
\bibitem{strom}A. Strominger, J. High Energy Phys. 02 (1998) 009.
\bibitem{sen}D. Birmingham, I. Sachs and S. Sen Phys. Lett. {\bf B413} (1997)281
.
\bibitem{lee}H. W. Lee, N. J. Kim and Y. S. Myung, hep-th/9803227.
\bibitem{hyun1} S. Hyun, Y. S. Song and J. H. Yee, Phys. Rev. {\bf D51} (1995) 1787.
\bibitem{btz2}M. Banados, M. Henneaux, C. Teitelboim and H. Zanelli, Phys. Rev. {\bf D48}
(1993) 1506.
\bibitem{avis} S. J. Avis, C. J. Isham, and D. Storey, Phys. Rev. {\bf D 18}(1978)3565.
\bibitem{bateman} A. Erdely, Higher Transedental Functions(Bateman Project), Vols I and II, (McGraw Hill, 1953).
\bibitem{unruh} W. Unruh, Phys. Rev. {\bf D14}(1976) 3251.
\bibitem{mals} J. Maldacena and A. Strominger, hep-th/9804085.
\bibitem{sachs} R. Emparan and I. Sachs, hep-th/9806122.
\end{references}
\end{document}